\documentclass[superscriptaddress,aps,preprintnumbers,amsmath,showpacs,amssymb,prd,nofootinbib,reprint]{revtex4-1}

\usepackage[dvipdfmx]{graphicx}
\usepackage{amsfonts}
\usepackage[mathscr]{eucal}
\usepackage{ascmac}
\usepackage{dcolumn}% Align table columns on decimal point
\usepackage{bm}% bold math
\usepackage[colorlinks=true,linkcolor=blue,citecolor=blue]{hyperref}
\usepackage{hyperref}
\usepackage{color}

\begin{document}

%%%%%%%%%%%%%%%%%%%%%%%%%%%%%%%%%%%%%%%%%%%

%\newcommand{\gtrsim}{ \mathop{}_{\textstyle \sim}^{\textstyle >} }
%\newcommand{\lesssim}{ \mathop{}_{\textstyle \sim}^{\textstyle <} }
\newcommand{\vev}[1]{ \left\langle {#1} \right\rangle }
\newcommand{\bra}[1]{ \langle {#1} | }
\newcommand{\ket}[1]{ | {#1} \rangle }
\newcommand{\eV}{ \ {\rm eV} }
\newcommand{\KeV}{ \ {\rm keV} }
\newcommand{\MeV}{\  {\rm MeV} }
\newcommand{\GeV}{\  {\rm GeV} }
\newcommand{\TeV}{\  {\rm TeV} }
\newcommand{\1}{\mbox{1}\hspace{-0.25em}\mbox{l}}
\newcommand{\Red}[1]{{\color{red} {#1}}}

\newcommand{\lmk}{\left(}  
\newcommand{\rmk}{\right)}
\newcommand{\lkk}{\left[}  
\newcommand{\rkk}{\right]}
\newcommand{\lhk}{\left \{ }  
\newcommand{\rhk}{\right \} }
\newcommand{\del}{\partial}  
\newcommand{\la}{\left\langle} 
\newcommand{\ra}{\right\rangle}
\newcommand{\half}{\frac{1}{2}}

\newcommand{\bea}{\begin{array}}
\newcommand{\eea}{\end{array}}
\newcommand{\beq}{\begin{eqnarray}}
\newcommand{\eeq}{\end{eqnarray}}

\newcommand{\dd}{\mathrm{d}}
\newcommand{\Mpl}{M_{\rm Pl}}
\newcommand{\mg}{m_{3/2}}
\newcommand{\abs}[1]{\left\vert {#1} \right\vert}
\newcommand{\mphi}{m_{\phi}}
\newcommand{\Hz}{\ {\rm Hz}}
\newcommand{\for}{\quad \text{for }}
\newcommand{\Min}{\text{Min}}
\newcommand{\Max}{\text{Max}}
\newcommand{\Kahler}{K\"{a}hler }
\newcommand{\cphi}{\varphi}
\newcommand{\Tr}{\text{Tr}}
\newcommand{\diag}{{\rm diag}}

\newcommand{\SUf}{SU(3)_{\rm f}}
\newcommand{\Upq}{U(1)_{\rm PQ}}
\newcommand{\Zpq}{Z^{\rm PQ}_3}
\newcommand{\Cpq}{C_{\rm PQ}}
\newcommand{\ubar}{u^c}
\newcommand{\dbar}{d^c}
\newcommand{\ebar}{e^c}
\newcommand{\nubar}{\nu^c}
\newcommand{\Ndw}{N_{\rm DW}}
\newcommand{\Fpq}{F_{\rm PQ}}
\newcommand{\fpq}{v_{\rm PQ}}
\newcommand{\Br}{{\rm Br}}
\newcommand{\Lag}{\mathcal{L}}
\newcommand{\Lqcd}{\Lambda_{\rm QCD}}
\newcommand{\const}{\text{const}}

\newcommand{\ji}{j_{\rm inf}} 
\newcommand{\jb}{j_{B-L}} 
\newcommand{\M}{M} 
\newcommand{\im}{{\rm Im} }
\newcommand{\re}{{\rm Re} }

%added by FT
\def\lrf#1#2{ \left(\frac{#1}{#2}\right)}
\def\lrfp#1#2#3{ \left(\frac{#1}{#2} \right)^{#3}}
\def\lrp#1#2{\left( #1 \right)^{#2}}
\def\REF#1{Ref.~\cite{#1}}
\def\SEC#1{Sec.~\ref{#1}}
\def\FIG#1{Fig.~\ref{#1}}
\def\EQ#1{Eq.~(\ref{#1})}
\def\EQS#1{Eqs.~(\ref{#1})}
\def\blue#1{\textcolor{blue}{#1}}
\def\red#1{\textcolor{blue}{#1}}

\newcommand{\fa}{f_{a}}
\newcommand{\Uh}{U(1)$_{\rm H}$}
\newcommand{\osc}{_{\rm osc}}

\newcommand{\mav}{\left. m_a^2 \right\vert_{T=0}}
\newcommand{\mat}{m_{a, {\rm QCD}}^2 (T)}
\newcommand{\mam}{m_{a, {\rm M}}^2 }
\def\eq#1{Eq.~(\ref{#1})}

\newcommand{\LQCD}{\Lambda_{\rm QCD}}

\newcommand{\UH}{U(1)$_H$ }

\newcommand{\EV}{ \ {\rm eV} }
\newcommand{\KEV}{ \ {\rm keV} }
\newcommand{\MEV}{\  {\rm MeV} }
\newcommand{\GEV}{\  {\rm GeV} }
\newcommand{\TEV}{\  {\rm TeV} }

%%%%%%%%%%%%%%%%%%%%%%%%%%%%%%%%%%%%%%%%%%%%%%%%%%%%%%%%%%%%%%%

\preprint{
TU-1022; \\
IPMU 16-0059
}

\title{
Diphoton Excess as a Hidden Monopole 
}

\author{
Masaki Yamada
}
\affiliation{Department of Physics, Tohoku University, 
Sendai, Miyagi 980-8578, Japan} 

\author{
Tsutomu T. Yanagida
}
\affiliation{Kavli IPMU (WPI), UTIAS, 
The University of Tokyo, 
Kashiwa, Chiba 277-8583, Japan}

\author{
Kazuya Yonekura
}
\affiliation{Kavli IPMU (WPI), UTIAS, 
The University of Tokyo, 
Kashiwa, Chiba 277-8583, Japan}

\date{\today}

\begin{abstract} 
We provide a theory with a monopole 
of a strongly-interacting hidden U(1) gauge symmetry 
that can explain the 750-GeV diphoton excess reported by ATLAS and CMS. 
The excess results from the resonance of monopole, 
which is produced via gluon fusion and decays into two photons. 
In the low energy, 
there are only mesons and a monopole in our model 
because any baryons cannot be gauge invariant in terms of strongly interacting Abelian symmetry. 
This is advantageous of our model 
because there is no unwanted relics around the BBN epoch.

\end{abstract}

\maketitle

%%%%%%%%%%%%%%%%%%%%%%%%%%%%%%%%%%%%%%%%%%%%%%%%%%%%%%%%%%%%%%%%
\section{Introduction
\label{sec:introduction}}
%%%%%%%%%%%%%%%%%%%%%%%%%%%%%%%%%%%%%%%%%%%%%%%%%%%%%%%%%%%%%%%%

The concepts of confinement and chiral symmetry breaking are long standing mystery 
in particle physics, which is not completely understood yet. 
It may seem to be natural that 
a non-abelian gauge theory is confined and develops quark condensation in low energy 
because its gauge coupling constant can be asymptotically free and may blow up at low energy. 
One might think that 
confinement does not occur in Abelian gauge theory, 
where the beta function of gauge coupling constant is positive in the presence of any charges of electrons. 
However, 
to understand the quark confinement in the color SU(3)$_c$ gauge theory, 
't Hooft conjectured that 
long-distance physics of non-Abelian gauge theory 
is dominated by its Abelian degrees of freedom~\cite{'tHooft:1981ht}. 
In fact, 
as Nambu showed, 
confinement can occur in an Abelian gauge theory 
where a scalar monopole as well as electrons and positrons are introduced~\cite{Nambu:1974zg}. 
Once the scalar monopole develops a condensation, 
each electron and positron pair is attached by a physical string 
and is confined by its tension.

In this paper we consider a phenomenological model of a U(1)$_H$ gauge theory with hidden electrons (quarks) and a monopole.
In fact, there really exist concrete models with the same qualitative features.
First of all, there exist conformal field theories (CFTs) 
which can be interpreted as U(1) gauge theories with electrons and 
monopoles \cite{Argyres:1995jj, Argyres:1995xn}. 
Given such abstract CFTs, we can deform the theory 
by introducing relevant operators whose scaling dimensions are less than 4.
In particular, in those U(1) theories, there exist a relevant operator
which can be interpreted as the electron masses [corresponding to our Eq.~(\ref{mass terms})],
and it was also discussed \cite{Bolognesi:2015wta} (see also \cite{Giacomelli:2014rna, Xie:2016hny, Buican:2016hnq}) that there exists a relevant operator 
which cause monopole condensation [corresponding to our Eq.~(\ref{potential})], 
leading to confinement and mesons.

In this paper, 
we provide a hidden Abelian gauge theory with a hidden monopole that explains 
diphoton excess at the energy scale of $750 \GEV$ 
reported by 
the ATLAS and CMS collaborators~\cite{ATLAS, CMS:2015dxe}.%
\footnote{
See Ref.~\cite{Yamada:2015waa} for another phenomenological application 
of monopole condensation. 
}
It has been discussed that 
its signal might be due to a resonance production of a composite particle 
that results in decay into diphoton (see, e.g., Ref.~\cite{Harigaya:2016eol} and references therein). 
In the previous works, 
they assumed that 
the composite state forms from additional particles 
which are charged under a hidden non-Abelian gauge symmetry. 
The strong dynamics of non-Abelian gauge interactions 
results in the confinement of additional particles 
and gives many composite states in the low energy, 
one of which has a mass of $750 \GEV$. 
In order to explain the diphoton excess and to avoid the constraint coming from the null results of 
LHC experiment at the energy scale of $8 \TEV$, 
the composite particle should be produced by a gluon fusion, 
which means that the additional particles have to be charged under the color SU(3)$_c$. 
In this case, however, 
there are problems in cosmology 
because 
baryon states, most of which are charged under the SU(3)$_c$, are stable or long-lived.%
\footnote{
When the baryon is neutral under the SM gauge interactions, 
it can be DM~\cite{Chiang:2015tqz}. 
}
The relics of such strongly interacting particles is severely constrained 
in cosmology, 
so that they should decay or be diluted efficiently before the BBN epoch. 
In this paper we point out that 
the cosmological problem can be naturally avoided 
when the composite states originate from a strong U(1)$_H$ gauge interaction. 
This is because there are no baryon states in low energy effective theory 
due to its U(1)$_H$ charge 
when the gauge symmetry is Abelian.

We first provide a different model which is easier to analyze 
and has some similarity with the model of U(1)$_H$ gauge symmetry. 
The former model consists of a singlet scalar field and 
extra quarks that are charged under hidden SU(N)$_H$ as well as SU(3)$_c$ $\times$ U(1)$_Y$. 
The Yukawa coupling between the scalar field and extra quarks 
can be as large as the gauge coupling of SU(N)$_H$, 
so that we can obtain a large cross section of diphoton signal for a large gauge coupling of SU(N)$_H$. 
This is similar to the above model with monopole in Abelian gauge theory, 
where the monopole interacts with extra quarks via strong U(1)$_H$ gauge interactions instead of Yukawa interaction. 
The advantageous of the theory with monopole 
is the absence of colored baryons, which are disastrous in cosmology if they are long lived or stable.

Then, in Sec.~\ref{model2}, 
we explain our model with a scalar monopole, 
where hidden quarks are confined by an Abelian gauge interaction 
due to a monopole condensation. 
As a result, 
there are a monopole and mesons in low energy 
and 
the former is responsible to the resonance of $750 \GEV$ at the LHC. 
Then we discuss cosmology 
and explain that the mesons are unstable and there is no unwanted relics around the BBN epoch. 
Finally, we conclude in Sec.~\ref{conclusions}.

\section{Model with SU(N)$_H$
\label{model}}

Let us consider a SU(N)$_H$ gauge theory with a singlet scalar field $\Phi$ 
and Weyl fermions $U$ and $\bar{U}$. 
The fields $U$ and $\bar{U}$ are charged under SU(N)$_H$ as well as 
the SM gauge symmetries as shown in Table~\ref{table0}. 
We call them as extra quarks 
because they are charged under the SU(3)$_c$. 
We introduce a Yukawa interaction such as 
\beq
 \mathcal{L} = y \Phi U \bar{U} + h.c., 
\eeq
and assume that $\Phi$ develops condensation such as $\la \Phi \ra \equiv v / \sqrt{2}$, which gives extra quarks an effective mass of $m_U \equiv y v / \sqrt{2}$.

%%%%%%%%%%%%%%%%%%%%%%%%%%%%%%%%%%%%%%%%%%%%%%%%%%%%%%%%%%%%%%%%
\begin{table}\begin{center}
\begin{tabular}{|p{1.0cm}|p{1.2cm}|p{1.2cm}|p{1.2cm}|p{1.2cm}|}
  \hline
  \rule[-5pt]{0pt}{15pt}
    & \hfil SU(3)$_c$ \hfil & \hfil SU(2)$_L$ \hfil & \hfil U(1)$_Y$ \hfil & \hfil SU(N)$_H$ \hfil  \\
  \hline
  \rule[-5pt]{0pt}{15pt}
  \hfil $U$ \hfil & \hfil $\Box$ \hfil & \hfil {\bf 1} \hfil & \hfil $q_Y$ \hfil & \hfil $\Box$ \hfil  \\
  \hline
  \rule[-5pt]{0pt}{15pt}
  \hfil $\bar{U}$ \hfil & \hfil $\bar{\Box}$ \hfil & \hfil {\bf 1} \hfil & \hfil $- q_Y$ \hfil & \hfil $\bar{\Box}$ \hfil  \\
\hline
\end{tabular}\end{center}
\caption{Charge assignment for extra matter fields in the model without monopole.
\label{table0}}
\end{table}
%%%%%%%%%%%%%%%%%%%%%%%%%%%%%%%%%%%%%%%%%%%%%%%%%%%%%%%%%%%%%%%%

Below the energy scale of quark mass $m_U$, 
we obtain the effective interaction of the phase direction of $\Phi$ such as 
\beq
 \mathcal{L} &=& 
 \frac{N \alpha_3}{8 \pi} \frac{ a }{v} G_{\mu \nu} \tilde{G}^{\mu \nu} 
 + \frac{3 N q_Y^2 \alpha_1}{4 \pi} \frac{ a }{v} B_{\mu \nu} \tilde{B}^{\mu \nu} 
\eeq
where we decompose the scalar field as $\Phi = (v +\cphi) e^{i a / v} / \sqrt{2}$. 
Thus, the decay rates of $a$ into SM gauge bosons are given as 
\beq
 \Gamma (a \to g g) 
 &\simeq&  \frac{N^2 \alpha_3^2 }{32 \pi^3 v^2} m_a^3
 \\
  \Br_{a \to \gamma \gamma} &=& 
 \frac{9 q_Y^4 \alpha^2}{2 \alpha_3^2} \simeq 3.5 \times 10^{-2} \times q_Y^4, 
\eeq
where we assume $\Br_{a \to g g} \simeq 1$ 
and use $\alpha_3 \simeq 0.09$ and $\alpha \simeq 1/126.5$ at the energy scale of $750 \GeV$. 
The mass of $a$, denoted as $m_a$, is independent of $m_U$ 
and assumed to be $750 \GEV$. 
The cross section of the process $\Gamma (pp \to a \to \gamma \gamma)$ 
at the center-of-mass energy $\sqrt{s} = 13 \TEV$ 
can be written as 
\beq
 \sigma (pp \to a \to \gamma \gamma) 
 &\simeq& 
 \mathcal{C}_{gg} 
 \frac{\Gamma (a \to g g)}{m_a s} \Br_{a \to \gamma \gamma} 
 \nonumber
 \\
 &\simeq& 
 3.5 {\rm \ fb} \ q_Y^4 y^{2} \lmk \frac{N}{3} \rmk^2 
 \lmk \frac{m_U}{1 \TEV} \rmk^{-2}, 
\eeq
where $\mathcal{C}_{gg} = (\pi^2/8) \int^1_0 \dd x_1 \int^1_0 \dd x_2 \delta 
(x_1 x_2 - m_\cphi^2/s) g(x_1) g(x_2)$ 
and $g(x)$ is the gluon parton distribution function. 
We use $\mathcal{C}_{gg} \simeq 2.1 \times 10^3$ from MSTW2008 NLO set 
at the scale of $m_a = 750 \GEV$~\cite{Martin:2009iq}. 
Here a large value of $y$ is crucial to make the quarks heavier than of order $1 \TEV$.

A large Yukawa coupling can be realized via the renormalization group running 
when the gauge coupling of SU(N)$_H$ is large. 
The RG equation is given by 
\beq
 && (16 \pi^2) \frac{\dd y}{\dd \log \mu} 
 \nonumber 
 \\
 &&= (3+3 N) y^3 - 
 \lmk 3 \frac{N^2 - 1}{N} g_H^2 + 8 g_3^2 + 6 q_Y^2 g_1^2 \rmk y, 
\eeq
so that the Yukawa coupling can be as large as $g_H$. 
Thus when the gauge coupling constant $g_H$ is large at the energy scale of order $m_a$, 
the Yukawa coupling can also be large and extra quark masses can be as large as of order $1 \TEV$. 

In this model, 
however, 
there is a stable baryon that is charged under the SU(3)$_c$, 
whose abundance is severely constrained in cosmology.%
\footnote{
The model might be safe if 
we could introduce 
a higher dimensional interaction between the extra quarks and SM particles~\cite{Harigaya:2016eol}. 
}
In the next subsection, 
we provide a theory that predicts no baryon and is safe cosmologically.

%%%%%%%%%%%%%%%%%%%%%%%%%%%%%%%%%%%%%%%%%%%%%%%%%%%%%%%%%%%%%%%%
\section{Model with U(1)$_H$ and a monopole 
\label{model2}}
%%%%%%%%%%%%%%%%%%%%%%%%%%%%%%%%%%%%%%%%%%%%%%%%%%%%%%%%%%%%%%%%

%%%%%%%%%%%%%%%%%%%%%%%%%%%%%%%%%%%%%%%%%%%%%%%%%%%%%%%%%%%%%%%%
\subsection{Model 
\label{model with monopole}}
%%%%%%%%%%%%%%%%%%%%%%%%%%%%%%%%%%%%%%%%%%%%%%%%%%%%%%%%%%%%%%%%

Now, we consider a hidden Abelian gauge theory with a scalar monopole $\phi$ 
and extra quarks 
$Q$, $\bar{Q}$, $U$, and $\bar{U}$. 
The charge assignment for the extra quarks are shown in Table~\ref{table1}. 
We denote the fine-structure constant for hidden electric charge as $\alpha_{H, e}$ 
($\equiv g_e^2 / 4 \pi$) 
and that for monopole charge as $\alpha_{H, m}$ ($\equiv g_m^2 / 4 \pi$), 
which satisfy Dirac quantization condition: 
\beq
 \alpha_{H, e} \alpha_{H, m} = \lmk \frac{n}{2} \rmk^2, 
\eeq
where $n$ is an integer. 
Hereafter we take $n=2$ as an example.

%%%%%%%%%%%%%%%%%%%%%%%%%%%%%%%%%%%%%%%%%%%%%%%%%%%%%%%%%%%%%%%%
\begin{table}\begin{center}
\begin{tabular}{|p{1.0cm}|p{1.2cm}|p{1.2cm}|p{1.8cm}|p{1.2cm}|}
  \hline
  \rule[-5pt]{0pt}{15pt}
    & \hfil SU(3)$_c$ \hfil & \hfil SU(2)$_L$ \hfil & \hfil U(1)$_Y$ \hfil & \hfil U(1)$_H$ \hfil  \\
  \hline
  \rule[-5pt]{0pt}{15pt}
  \hfil $Q$ \hfil & \hfil $\Box$ \hfil & \hfil $\Box$ \hfil & \hfil $q_Y$ \hfil & \hfil 1 \hfil  \\
  \hline
  \rule[-5pt]{0pt}{15pt}
  \hfil $U$ \hfil & \hfil $\Box$ \hfil & \hfil {\bf 1} \hfil & \hfil $q_Y-1/2$ \hfil & \hfil 1 \hfil  \\
  \hline
  \rule[-5pt]{0pt}{15pt}
  \hfil $\bar{Q}$ \hfil & \hfil $\bar{\Box}$ \hfil & \hfil $\bar{\Box}$ \hfil & \hfil $-q_Y$ \hfil & \hfil -1 \hfil  \\
  \hline
  \rule[-5pt]{0pt}{15pt}
  \hfil $\bar{U}$ \hfil & \hfil $\bar{\Box}$ \hfil & \hfil {\bf 1} \hfil & \hfil $- (q_Y-1/2)$ \hfil & \hfil -1 \hfil  \\
\hline
\end{tabular}\end{center}
\caption{Charge assignment for extra matter fields in the model with monopole.
\label{table1}}
\end{table}
%%%%%%%%%%%%%%%%%%%%%%%%%%%%%%%%%%%%%%%%%%%%%%%%%%%%%%%%%%%%%%%%

\begin{figure}[t]
\centering 
\includegraphics[width=.40\textwidth, bb=0 0 360 351]{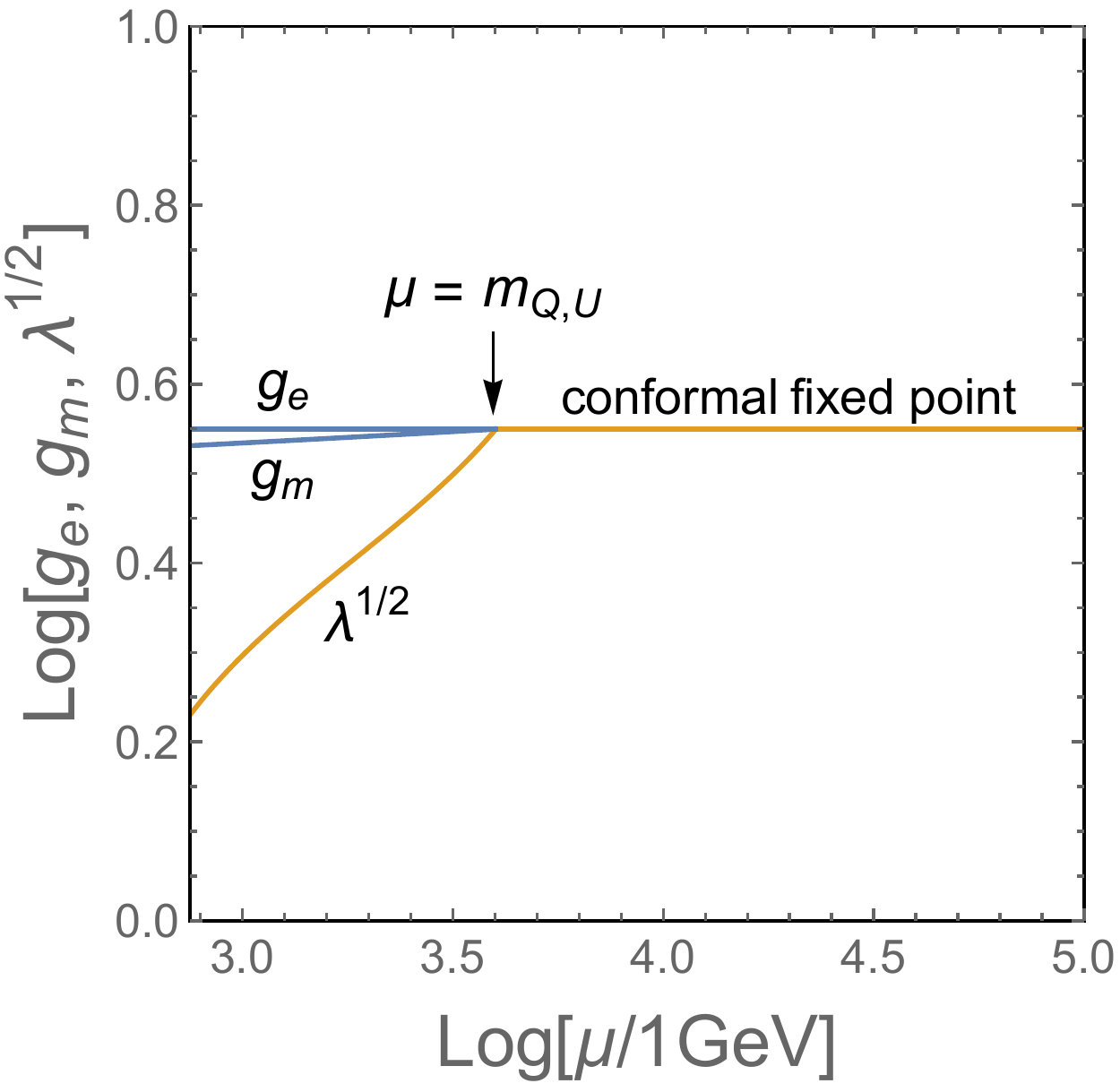} 
\caption{
Schematic plot of RG flow for gauge couplings $g_e$ and $g_m$ and 
monopole quartic coupling $\lambda$. 
The masses of extra quarks are taken to be $m_Q = m_U = 4 \TEV$. 
The dimensionless parameters are assumed to be $g_m = g_e = \lambda^{1/2} = (4 \pi)^{1/2}$ 
at the UV fixed point, where the energy scale is higher than the mass of extra quarks. 
}
  \label{fig0}
\end{figure}

In the presence of the hidden magnetic monopole as well as the hidden electrically charged particles, 
the theory may be conformal 
and the coupling constants are of order unity at the UV fixed point~\cite{Argyres:1995jj, Argyres:1995xn}. 
Once we add mass terms for extra quarks such as 
\beq
 - \mathcal{L} = m_Q Q \bar{Q} + m_U U \bar{U}, 
 \label{mass terms}
\eeq
then
the hidden electrically charged particles are decoupled 
below these mass scale. 
The U(1)$_H$ theory contains only a scalar monopole after the extra quarks decouple, 
which means that the low energy theory is equivalent to a scalar QED 
due to the electromagnetic duality of U(1)$_H$. 
When we write the potential of scalar monopole such as 
\beq
  V(\phi) = - \mu^2 \abs{\phi}^2 + \lambda \abs{\phi}^4, 
  \label{potential}
\eeq
the renormalization group (RG) equations can be written as 
\beq
 (16 \pi^2) \frac{\dd g_m}{\dd \ln \mu} 
 &=& \frac{g_m^3}{3} 
 \\
 (16 \pi^2) \frac{\dd \lambda}{\dd \ln \mu} 
 &=&  20 \lambda^2 - 12 g_m^2 \lambda + 6 g_m^4, 
 \label{RG}
\eeq
within one-loop order. 
The RG group flow is shown in Fig.~\ref{fig0}, 
where we assume $g_m = g_e = \lambda^{1/2} = (4 \pi)^{1/2}$ at the UV fixed point. 
Note that the couplings are larger than unity, so that 
the above perturbative calculation cannot be trusted and the figure should be regarded as a schematic plot. 
The theory is at the UV fixed point in the energy scale higher than $m_{Q, U}$ 
and the coupling constants run below that scale. 
The magnetic coupling becomes smaller 
in lower energy scale 
while the electric coupling becomes larger due to the Dirac's quantization condition. 
Note that 
in this theory 
there are only two free parameters: quark mass scale $m_{Q, U}$ and monopole mass parameter $\mu$. 
The other parameters, such as electric and magnetic couplings 
and monopole quartic coupling, 
can be determined in principle by the RG running from 
the UV fixed point. 
However, 
we do not know the values of these parameters at the UV fixed point, 
so that 
we just expect that 
these couplings at the UV fixed point are of order unity. 
The couplings at the energy scale of $750 \GEV$ should be determined by 
solving renormalization group equations between the quark mass scale and $750 \GEV$. 
The RG equations of Eq.~(\ref{RG}) imply that 
the quartic coupling of monopole $\lambda$ 
may be smaller than the hidden magnetic coupling $g_m$ in low energy (see Fig.~\ref{fig0}).

%%%%%%%%%%%%%%%%%%%%%%%%%%%%%%%%%%%%%%%%%%%%%
%%%%%%%%%%%%%%%%%%%%%%%%%%%%%%%%%%%%%%%%%%%%%

At the minimum of the potential, 
the monopole develops a condensation such as $\sqrt{2} \la \abs{\phi} \ra = \mu / \sqrt{\lambda}$ 
($\equiv v$) 
and the mass of its radial component, which we denote as $\cphi$ ($\equiv \sqrt{2} \abs{\phi}$), 
is given by $m_\cphi = \sqrt{2} \mu$. 
The hidden \UH gauge boson acquires a mass of $m_v \equiv g_m v$, 
which we assume to be larger than $m_\cphi$. 
After the monopole acquires the VEV, 
extra quarks are attached by strings via the Meisner effect 
and are 
confined by the tension of the string~\cite{Nambu:1974zg}. 
Its tension $\mu_s$ determines the dynamical scale and is given as 
\beq
 \mu_s = 2 \pi \alpha_{H, e} \alpha_{H, m} v^2 \log \lmk \frac{m_\cphi^2}{m_v^2} + 1 \rmk, 
\eeq
which is almost independent of $\alpha_{H, e}$ and $\alpha_{H, m}$ due to the Dirac quantization condition. 
We assume that the extra quark masses $m_{Q, U}$ are larger 
than the confinement scale, 
so that 
there 
is only the radial component of monopole 
below the confinement scale. 
We identify the monopole as a particle with a mass of $750 \GEV$ 
that is responsible for the diphoton resonance.

%%%%%%%%%%%%%%%%%%%%%%%%%%%%%%%%%%%%%%%%%%%%%%%%%%%%%%%%%%%%%%%%
\subsection{Collider signals
\label{diphoton}}
%%%%%%%%%%%%%%%%%%%%%%%%%%%%%%%%%%%%%%%%%%%%%%%%%%%%%%%%%%%%%%%%

The monopole and SM gauge fields are coupled 
via the heavy quarks. 
Therefore the couplings is inversely proportional to the typical mass scale of $Q$ and $U$. 
In the naive dimensional analysis~\cite{Manohar:1983md}, 
the couplings between the monopole and SM gauge fields are given by 
\beq
 \mathcal{L} &=& 
 c_3 \frac{g_3^2}{16 \pi^2} \frac{(4\pi)^2 \abs{\phi}^2}{m_{Q, U}^2} G_{\mu \nu} G^{\mu \nu} 
\nonumber \\
 &&+ c_2 \frac{g_2^2}{16 \pi^2} \frac{(4\pi)^2 \abs{\phi}^2}{m_{Q, U}^2} W_{\mu \nu} W^{\mu \nu} 
\nonumber \\ 
 &&+ c_1 \frac{g_1^2}{16 \pi^2} \frac{(4\pi)^2 \abs{\phi}^2}{m_{Q, U}^2} B_{\mu \nu} B^{\mu \nu}, 
\eeq
where $c_i$ ($i = 1, 2, 3$) are unknown $O(1)$ constants and $m_{Q, U}$ is a typical mass parameter 
of order $m_Q$ and $m_U$. 
Then, 
using $\abs{\phi}^2 = v^2/2 + v \cphi + \dots$, 
we obtain 
the decay rates of $\cphi$ into SM gauge bosons such as 
\beq
 \Gamma (\cphi \to g g) 
 &\simeq& \frac{8  c_3^2 \alpha_c^2 (4 \pi v)^2}{4 \pi m_{Q, U}^4} m_\cphi^3
 \\
  \Br_{\cphi \to \gamma \gamma} &=& 
 \frac{(c_1 +c_2)^2 e^4}{8 c_3^2 g_3^4}, 
 \\
 \Br_{\cphi \to W^+ W^-} &=& 
 \frac{c_2^2 e^4}{4 s_W^2  c_3^2 g_3^4}, 
 \\
 \Br_{\cphi \to Z Z} &=& 
 \frac{(c_1 t_W^2 +c_2/t_W^2)^2 e^4}{8 c_3^2 g_3^4}, 
 \\
 \Br_{\cphi \to Z \gamma} &=& 
  \frac{(c_1 / t_W -  c_2 t_W)^2 e^4}{4 c_3^2 g_3^4}, 
\eeq
where we assume $\Br_{\cphi \to g g} \simeq 1$ 
and define $s_W \equiv \sin \theta_W$ and $t_W \equiv \tan \theta_W$. 
The cross section of the process $\Gamma (pp \to \cphi \to \gamma \gamma)$ 
can be written as 
\beq
 \sigma (pp \to \cphi \to \gamma \gamma) 
 &\simeq& 
 \mathcal{C}_{gg} 
 \frac{\Gamma (\cphi \to g g)}{m_\cphi s} \Br_{\cphi \to \gamma \gamma} 
 \nonumber
 \\
 &\simeq& 
 3.8 {\rm \ fb} 
 (c_1+c_2)^2 
 \lmk \frac{\lambda}{10} \rmk^{-1}
 \lmk \frac{m_{Q, U}}{2 \TEV} \rmk^{-4}, 
\nonumber \\
\eeq
where we use $\mathcal{C}_{gg} \simeq 2.1 \times 10^3$, $m_\cphi = 750 \GEV$, 
and $\alpha_c \simeq 0.09$ 
for $\sqrt{s} = 13 \TEV$. 
We find that the masses of extra quarks can be larger than $\mathcal{O}(1) \TEV$ 
for $\lambda = \mathcal{O}(10)$. 

The hidden vector boson mass $m_v$ is related to the value of $\lambda$ 
such as $m_v = g_m v = g_m m_\cphi / \sqrt{2 \lambda}$. 
Assuming that 
$g_m \approx \sqrt{4 \pi}$ and $\lambda \simeq 3-10$, 
which is expected from the RG running (see Fig.~\ref{fig0}), 
we estimate $m_v \approx 0.6 - 1.1 \TEV$. 
Note that 
the hidden gauge boson acquires an effective mass 
by the monopole condensation, 
so that the hidden gauge coupling does not mix with the electroweak coupling. 
Thus the hidden gauge boson, which we denote by $Z'$, can be produced by collider experiments 
only via loop effects. 
A model that predicts similar signals has been investigated in Ref.~\cite{Alwall:2012np}. 
They focused on $g g \to Z' g$ associated with $Z' \to g g^* \to g t \bar{t}$, 
where $g$ and $g^*$ represent on-shell and off-shell gluons, respectively. 
Since the scattering cross section is suppressed by the masses of extra quarks, 
which is of order $2-4 \TEV$, 
the signal is much below background signals. 
A process $p \bar{p} \to Z + {\rm jets}$ is also predicted in our model. 
However, its background signal cross section is as large as $10^{4-5} {\rm \ pb}$~\cite{Aad:2016naf}, 
so that we cannot obtain any signals from this process.

Here we comment on a consequence from sequestering property of conformal field theory~\cite{Argyres:1995jj, Argyres:1995xn}. 
One may naively expect that 
the decay rate into diphoton is roughly proportional to $q_Y^4$ 
for $q_Y \gg 1$ 
because its process is mediated by extra quarks with U(1)$_Y$ charge of order $q_Y$. 
However, 
this may not be the case in the conformal Abelian gauge theory. 
First, note that 
the hypercharge of $Q$, $\bar{Q}^\dagger$, $U$, and $\bar{U}^\dagger$ 
can be shifted by $-q_Y$ 
by redefinition of U(1)$_H$ gauge field 
and be rewritten by a kinetic mixing term between U(1)$_Y$ and \UH 
($\propto q_Y B_{\mu \nu} F^{\mu \nu}$, 
where $F_{\mu \nu}$ is the field strength of \UH). 
When the \UH gauge theory is conformal, 
its gauge field strength $F_{\mu \nu}$ has an anomalous dimension larger than $2$. 
This implies that 
the kinetic mixing term $B_{\mu \nu} F^{\mu \nu}$ is an irrelevant operator 
and is suppressed at low energy. 
As a result, 
if the hypercharges of $Q$, $\bar{Q}^\dagger$, $U$, and $\bar{U}^\dagger$ 
were identical, 
their hypercharge would be suppressed in the low energy effective theory. 
This is the reason why 
we do not expect that the decay rate of monopole into diphoton is proportional to $q_Y^4$. 
Since their hypercharges are not identical in our model, 
we expect nonzero decay rate of monopole into diphoton. 
Note that if the kinetic mixing term is not suppressed at low energy, 
we do not need to introduce $Q$ and $\bar{Q}$ to explain the diphoton signal.

%%%%%%%%%%%%%%%%%%%%%%%%%%%%%%%%%%%%%%%%%%%%%%%%%%%%%%%%%%%%%%%%
\subsection{Mesons and cosmology 
\label{cosmology}}
%%%%%%%%%%%%%%%%%%%%%%%%%%%%%%%%%%%%%%%%%%%%%%%%%%%%%%%%%%%%%%%%

In the low energy effective theory, 
there are mesons as well as monopoles. 
Their SM gauge charges are listed in Table~\ref{table2}. 
As we discussed in the previous section, 
the mass of the mesons are as heavy as $2 m_{Q, U} \approx 4 \TEV$, 
so that we may not be able to produce them at the LHC. 
In any case, 
we should check that they do not affect the standard cosmological scenario, 
such as the BBN theory.

%%%%%%%%%%%%%%%%%%%%%%%%%%%%%%%%%%%%%%%%%%%%%%%%%%%%%%%%%%%%%%%%
\begin{table}\begin{center}
\begin{tabular}{|p{1.0cm}|p{1.2cm}|p{1.2cm}|p{1.2cm}|}
  \hline
  \rule[-5pt]{0pt}{15pt}
    & \hfil SU(3)$_c$ \hfil & \hfil SU(2)$_L$ \hfil & \hfil U(1)$_Y$ \hfil  \\
  \hline
  \rule[-5pt]{0pt}{15pt}
  \hfil $\psi_Q$ \hfil & \hfil {\bf Adj} \hfil & \hfil {\bf Adj} \hfil & \hfil $0$ \hfil \\
  \hline
  \rule[-5pt]{0pt}{15pt}
  \hfil $\psi_Q'$ \hfil & \hfil {\bf Adj} \hfil & \hfil {\bf 1} \hfil & \hfil $0$ \hfil \\
  \hline
  \rule[-5pt]{0pt}{15pt}
  \hfil $\psi_Q''$ \hfil & \hfil {\bf 1} \hfil & \hfil {\bf Adj} \hfil & \hfil $0$ \hfil \\
  \hline
  \rule[-5pt]{0pt}{15pt}
  \hfil $\pi^\pm$ \hfil & \hfil {\bf Adj} \hfil & \hfil $\Box$ \hfil & \hfil $\pm 1/2$ \hfil \\
  \hline
  \rule[-5pt]{0pt}{15pt}
  \hfil $\pi'^\pm$ \hfil & \hfil {\bf 1} \hfil & \hfil $\Box$ \hfil & \hfil $\pm 1/2$ \hfil \\
  \hline
  \rule[-5pt]{0pt}{15pt}
  \hfil $\psi_U$ \hfil & \hfil {\bf Adj} \hfil & \hfil {\bf 1} \hfil & \hfil $0$ \hfil \\
  \hline
  \rule[-5pt]{0pt}{15pt}
  \hfil $\eta, \eta'$ \hfil & \hfil {\bf 1} \hfil & \hfil {\bf 1} \hfil & \hfil $0$ \hfil \\
\hline
\end{tabular}\end{center}
\caption{SM gauge charges of extra mesons.
\label{table2}}
\end{table}
%%%%%%%%%%%%%%%%%%%%%%%%%%%%%%%%%%%%%%%%%%%%%%%%%%%%%%%%%%%%%%%%

In the low energy, 
we can write the following operators allowed by symmetry: 
\beq
 &&\mathcal{L} 
= 
 \frac{\kappa_1 g_2 g_3}{32 \pi^2 v} \psi_Q^{a \alpha} \epsilon^{\mu \nu \rho \sigma} 
 G_{\mu \nu}^a W_{\rho \sigma}^\alpha 
 + 
 \frac{\kappa_2 g_3^2}{32 \pi^2 v} d^{abc} \psi'^a_Q \epsilon^{\mu \nu \rho \sigma} 
 G_{\mu \nu}^b G_{\rho \sigma}^c 
 \nonumber \\
 &&+
 \frac{\kappa_3 g_1 g_3}{32 \pi^2 v} \psi'^a_Q \epsilon^{\mu \nu \rho \sigma} 
 G_{\mu \nu}^a B_{\rho \sigma} 
+
 \frac{\kappa_4 g_1 g_2}{32 \pi^2 v} \psi''^\alpha_Q \epsilon^{\mu \nu \rho \sigma} 
 W_{\mu \nu}^\alpha B_{\rho \sigma} 
 \nonumber \\
 &&+
 \frac{\kappa_5 g_1 g_3}{32 \pi^2 v} \psi_U^a \epsilon^{\mu \nu \rho \sigma} 
 G_{\mu \nu}^a B_{\rho \sigma} 
 \nonumber \\
 &&+ 
 \frac{\kappa_6 g_2^2}{32 \pi^2 v} \eta \epsilon^{\mu \nu \rho \sigma} 
 W_{\mu \nu}^\alpha W_{\rho \sigma}^\alpha 
+
 \frac{\kappa_7 g_1^2}{32 \pi^2 v} \eta \epsilon^{\mu \nu \rho \sigma} 
 B_{\mu \nu} B_{\rho \sigma} 
\nonumber \\
 &&+ \frac{\kappa_8 g_3^2}{32 \pi^2 v} \eta' \epsilon^{\mu \nu \rho \sigma} 
 G_{\mu \nu}^a G_{\rho \sigma}^a 
 + \frac{\kappa_9 g_2^2}{32 \pi^2 v} \eta' \epsilon^{\mu \nu \rho \sigma} 
 W_{\mu \nu}^\alpha W_{\rho \sigma}^\alpha 
 \nonumber \\
&& +
 \frac{ \kappa_{10} g_1^2}{32 \pi^2 v} \eta' \epsilon^{\mu \nu \rho \sigma} 
 B_{\mu \nu} B_{\rho \sigma}, 
\eeq
where $\kappa_i$ are $\mathcal{O}(1)$ factors 
and $d^{abc} \equiv 2 {\rm Tr} [ t^a \{ t^b, t^c \} ]$ 
with $t^a$ being half of the Gell-Mann matrices. 
Thus $\psi_Q$, $\psi'_Q$, $\psi''_Q$, $\psi_U$, $\eta$, and $\eta'$ can decay into the SM gauge bosons. 
Note that the mass of $\eta'$ is the same order with that of the other mesons 
because the dynamical scale $v$ 
is much smaller than the typical mass scale of extra quarks $m_Q$ and $m_U$. 
In order to make the other mesons ($\pi^\pm$ and $\pi'^\pm$) decay, 
we introduce interactions of 
\beq
 \mathcal{L}_{\rm int} 
 = 
 y Q \bar{U} H 
+ h.c., 
\eeq
where $y$ is a Yukawa coupling constant. 
This interaction allows $\pi^\pm$ and $\pi'^\pm$ to decay into SM particles 
so fast 
that we can avoid the BBN constraint.

%%%%%%%%%%%%%%%%%%%%%%%%%%%%%%%%%%%%%%%%%%%%%%%%%%%%%%%%%%%%%%%%
\section{Discussion and conclusions
\label{conclusions}}
%%%%%%%%%%%%%%%%%%%%%%%%%%%%%%%%%%%%%%%%%%%%%%%%%%%%%%%%%%%%%%%%

We have provided a simple model that explains the diphoton excess reported by ATLAS and CMS 
and predicts no unwanted relics in the Universe. 
First we explain a model with non-Abelian gauge symmetry to 
illustrate our mechanism, where extra quarks can be as heavy as $\mathcal{O}(1) \TEV$. 
Then we discuss our model based on a confinement in Abelian gauge theory, 
which can be realized by a monopole condensation at an intermediate scale. 
The diphoton excess results from the resonance production of 750-GeV monopole 
by gluon fusion and its subsequent decay into diphoton. 
We predict mesons with the masses of order $5 \TEV$, 
which decay fast and do not spoil the success of the BBN theory in cosmology. 
We also predict a massive hidden gauge boson with mass about $1 \TEV$. 
Since it acquires an effective mass by the monopole condensation, 
its production process is different from the ordinary $Z'$. 
It may be challenging to search its signals in LHC.

Finally, we comment on another interesting possibility for application of monopole condensation in cosmology and phenomenology. 
As discussed in the final paragraph in Sec.~\ref{diphoton}, 
the kinetic mixing between U(1)$_Y$ and hidden U(1)$_H$ is suppressed in low energy 
when the hidden U(1)$_H$ is conformal due to the presence of monopole as well as electrons. 
The suppression depends on the (unknown) anomalous dimension of field strength of U(1)$_H$, 
so that the kinetic mixing may be a nonzero small value. 
This provides a simple mechanism to suppress the kinetic mixing of Abelian gauge theories, 
which is severely constrained by many experiments. 

In the literature, strongly interacting massive particles (SIMPs) are well motivated as DM 
in light of a solution to a tension between the cold DM model and astrophysical observations (see e.g., Refs.~\cite{Spergel:1999mh, deBlok:2009sp, BoylanKolchin:2011de}). 
In Ref.~\cite{Hochberg:2014kqa}, 
they considered strongly-interacting hidden SU(N) gauge theory 
and identified pions in low energy effective theory as SIMPs. 
Their relic abundance is determined by $3 \to 2$ scattering 
and can explain the observed one around a parameter space consistent with astrophysical observations~\cite{Hochberg:2014dra}. 
However, we need interactions between the pions and SM sector 
so that the energy of pions can be reduced in order not to be hot DM. 
In Ref.~\cite{Hochberg:2015vrg}, they introduced an additional U(1)$_H$ gauge symmetry 
and assume a kinetic mixing between the U(1)$_H$ and U(1)$_Y$. 
Here, we can consider a simpler model 
where the above hidden SU(N) is replaced by our hidden U(1)$_H$. 
The hidden electrons are confined by the monopole condensation, leading
the electron and anti-electron chiral condensation. 
The kinetic mixing 
between U(1)$_H$ and U(1)$_Y$ 
can be naturally small as discussed above and allows pions to reduce its energy 
without affecting their relic abundance. 
In addition, 
there is no unwanted baryon in this theory as discussed in the main part of this paper. 
A detailed study will be presented elsewhere~\cite{future work}. 

\vspace{1cm}

%---------------SECTION------------------%
%
\section*{Acknowledgments}
T.~T.~Y thanks Chengcheng~Han for useful comments on collider signatures. 
This work is supported by Grant-in-Aid for Scientific Research 
from the Ministry of Education, Science, Sports, and Culture
(MEXT), Japan, 
No. 26104009 and No. 26287039 (T.T.Y), 
World Premier International Research Center Initiative
(WPI Initiative), MEXT, Japan, 
and the JSPS Research Fellowships for Young Scientists (M.Y.). 
%
%---------------SECTION------------------%

\vspace{1cm}

%%%%%%%%%%%%%%%%%%%%%%%%%%%%%%%%%%%%%%%%%%%%%%%%%%

%%%%%%%%%%%%%%%%%%%%%%%%%%%%%%%%%%%%%%%%%%%%%%%%%%

\end{document}